\documentclass[showpacs,twocolumn,prb]{revtex4}
\usepackage{graphicx}
\usepackage{amsmath}
\newcommand{\be}{\begin{equation}}
\newcommand{\ee}{\end{equation}}
\newcommand{\bea}{\begin{eqnarray}}
\newcommand{\eea}{\end{eqnarray}}
\newcommand{\bwt}{\begin{widetext}}
\newcommand{\ewt}{\end{widetext}}

\begin{document}
\title{The harmonic oscillator in quantum mechanics: A third way}
\author{F. Marsiglio}
\affiliation{Department of Physics, University of Alberta, Edmonton, Alberta, Canada, T6G~2J1}

\begin{abstract}
Courses on undergraduate quantum mechanics usually focus on solutions of the Schr\"odinger equation for
several simple one-dimensional examples. When the notion of a Hilbert space is introduced only academic
examples are used, such as the
matrix representation of Dirac's raising and lowering operators or the angular momentum operators. We
introduce some of the same one-dimensional examples as matrix diagonalization problems, with a basis that
consists of the infinite set of square well eigenfunctions. Undergraduate students are well equipped to handle
such problems in familiar contexts. We pay special attention to the one-dimensional harmonic oscillator. This
paper should equip students to obtain the low lying bound states of any one-dimensional short range potential.
\end{abstract}

\date{\today}
\maketitle

\section{introduction}

A standard undergraduate text in quantum mechanics \cite{griffiths} devotes most of its attention to
analytical solutions of the Schr\"odinger equation. Some texts\cite{feagin,thaller} coordinate problem solving
in quantum mechanics with numerical solutions. However, the focus remains the solution of the Schr\"odinger
equation.

A typical introductory course also includes a treatment of the more formal part of quantum mechanics, with
topics such as vector spaces and the matrix representation. This part is usually more abstract. Examples that
are often used in this part are two state systems and matrix representations of angular momentum.
Alternatively, an actual (say, $3 \times3$) matrix is given, and students are asked to use matrix algebra to
mimic a solution of a quantum mechanical problem. These examples often strike students as artificial. The main
part of such a course appears to students to be the solution of various differential equations, and the formal
part of the course often appears to be superfluous.

Numerical solutions of the time-independent Schr\"odinger equation have been discussed in various contexts
(see, for example, Ref.~\onlinecite{kinderman90}). These solutions utilize numerical techniques for solving a
differential equation. Even in the rare instance that an actual basis set is used,\cite{belloni08} it is used
to determine the time evolution for the time-dependent Schr\"odinger equation. Many quantum mechanical
research problems that are amenable to solution (for example, the behavior of electrons on a small lattice)
are solved using matrix mechanics.\cite{dagotto94} The purpose of this note is to bring attention to
a wide variety of time-independent problems that can be tackled through matrix mechanics at an introductory
level.

In the following section we set up the infinite square well  problem. The solution to this problem constitutes
the (infinite) basis set with which we can tackle any single particle potential problem that can be embedded
in the infinite square well. In Sec.~III we illustrate the matrix solutions for the harmonic oscillator
potential. This example is chosen because the students will have already encountered it using the algebraic
method (series solution) and Dirac's operator method. The former method tends to leave students struggling
with the mathematics, and the latter method leaves many students awestruck. The method proposed here uses
matrix algebra, which is a more familiar mathematical method.

We also illustrate wave functions, which can be readily calculated from the eigenvector solutions. We have
been unable to find problems of this nature in the literature. One reference illustrates a solution to the
three-dimensional harmonic oscillator bounded by two impenetrable walls, but does so following the standard
route of differential equations, though in cylindrical coordinates.\cite{marin88}

The matrix methodology discussed here is at the heart of many applications in physics. For example, vibrations
in solids\cite{dove93} are treated using springs that obey harmonic oscillator potentials. Bose condensation
in alkali gases\cite{anderson95} utilizes a harmonic trap, and the conventional theory of superconductivity
uses electron-ion interactions with harmonic oscillators.\cite{schrieffer84} Calculations of a single electron
interacting with vibrating ions (that is, a polaron) that use matrix diagonalization can be found in
Ref.~\onlinecite{marsiglio93}.

The convergence criterion for the harmonic oscillator potential is related to energy scales, and the number of
basis states required to accurately represent these solutions is dictated by the energies of the basis states.
In Sec.~IV we briefly discuss some other familiar interactions, such as the
finite square well and the delta function potential. In some of these cases length scales also play an
important role, which leads to a discussion of the convergence criterion connected with capturing the short
length scales in the potential.

This paper is meant to be sufficiently self-contained and should serve to enhance the undergraduate curriculum
in quantum mechanics, and equip undergraduates to solve a variety of useful problems.

\section{the formalism}

A typical introductory course begins with one-dimensional problems, including the free particle, the infinite
square well, and the harmonic oscillator potential. The free particle has certain pathologies associated with
the normalization of the stationary states. For this reason we prefer to discuss the infinite square well, in
which a particle is confined to the space enclosed by the walls of the well. The harmonic oscillator is also
amenable to analytical solution. Unlike the infinite square well the solutions (Hermite polynomials enveloped
by Gaussians) are unfamiliar to the novice student, and, as far as a typical student is concerned, might as
well be numerical.

For these reasons we focus on the infinite square well, which is defined as \be V_{\text inf}(x) = \begin{cases} 0 &
\text{if $0 < x < a$,} \\ \infty & \text{otherwise.}
\end{cases}
\label{infinite_square_well_potential} \ee The Hamiltonian is given by
\begin{equation}
H_0 = -{\hbar^2 \over 2m} {d^2 \over dx^2} + V_{\text inf}(x), \label{ham0}
\end{equation}
where $m$ is the mass of the particle. The eigenstates are well known: \be \psi_n(x) = \begin{cases}
\sqrt{\dfrac{2}{a}} \sin{\Big( \dfrac{n \pi x}{a} \Big)} & \text{if $0 < x < a$,} \\ 0 & \text{otherwise},
\end{cases}
\label{infinite_square_well_wavefunction} \ee with eigenvalues,
\begin{equation}
E_n^{(0)} = {n^2 \pi^2 \hbar^2 \over 2 m a^2} \equiv n^2E_1^{(0)}. \label{infinite_square_well_energies}
\end{equation}
The quantum number $n = 1,2,3, \ldots$ takes on a positive integer value. Students are generally unaware of
perturbation theory at this stage, but reasons for the superscript $(0)$ can be readily explained.

Although we have presented the well-known solutions for the infinite square well problem, the emphasis in the
preceding paragraphs should be on the formulation and presentation of a convenient and familiar set of basis
functions. We are now prepared to tackle a variety of problems, represented by $H = H_0 + V(x)$, where $H_0$
is as in Eq.~\eqref{ham0}, and $V(x)$ is any potential in the domain $0<x<a$. We proceed as in most textbooks,
except now we have the specific basis set Eq.~(\ref{infinite_square_well_wavefunction}) in mind.

We start with the Schrodinger equation in ket notation: \be (H_0 + V) | \psi \rangle = E|\psi \rangle.
\label{a1} \ee If we use the general expansion of a wave function in terms of a complete set of basis states,
\be | \psi \rangle = \sum_{m=1}^\infty c_m |\psi_m \rangle, \label{a2} \ee we obtain \be \sum_{m=1}^\infty c_m
(H_0 + V)| \psi_m \rangle = E\sum_{m=1}^\infty c_m |\psi_m \rangle. \label{a3} \ee By taking the inner product
of \eqref{a3} with the bra $\langle \psi_n|$, we obtain the matrix equation: \be \sum_{m=1}^\infty H_{nm} c_m
= E c_n, \label{a4} \ee where
%
\begin{eqnarray}
& & H_{nm}  =  \langle \psi_n|(H_0 + V)|\psi_m \rangle =  \nonumber \\
&&\delta_{nm}E_n^{(0)}+ {2 \over a}\int_0^a dx  \sin{\bigl( {n \pi x \over a} \bigr)}
V(x)  \sin{\bigl( {m \pi x \over a} \bigr)}
\label{ham_matrix}
\end{eqnarray}
is the Hamiltonian matrix; in the second term we have used the eigenstates of $H_0$ as the basis set. As
usual, $\delta_{nm}$ is the Kronecker delta function.

Equation~(\ref{a4}) along with Eq.~(\ref{ham_matrix}) allows us to study a variety of interesting problems,
albeit in the space determined by the infinite square well. Note that for numerical solutions of
Eq.~(\ref{a4}) we need to truncate the sum and hence the number of eigenstates used to describe the solutions.
We denote the maximum basis set number by $N$.

\section{the harmonic oscillator}

We imbed the harmonic oscillator potential, $V_{\rm HO} = m \omega^2x^2/2$, in the infinite square well.
In the following all units of distance will be in terms of the square well width $a$, and all units of energy
will be in terms of the (unperturbed) ground state, $E_1^{(0)}$. We will use lower case letters to denote
dimensionless energies. The potential $V_{\rm HO}$ can
be written in terms of the infinite square well length and energy scales as \be v_{\rm HO} = {V_{\rm HO} \over
E_1^{(0)}} = {\pi^2 \over 4} \bigg({\hbar \omega \over E_1^{(0)}} \bigg)^2 \Big({x \over a} - {1 \over
2}\Big)^2, \label{ho_pot} \ee so that the dimensionless
parameter ${\hbar \omega/E_1^{(0)}}$ determines the stiffness of the harmonic oscillator potential. We expect
that for low energy states (say, the ground state), the solution should be identical to that of the harmonic
potential alone, because the wave function will be sufficiently restricted to the central region of the
harmonic oscillator potential so that it will not ``feel'' the walls of the infinite square well. High energy
states will not be well described by the harmonic oscillator results, because they will be primarily governed
by the infinite square well.

We first use Eq.~(\ref{ham_matrix}) with the potential given by Eq.~(\ref{ho_pot}). The result is 
\bea
&&h_{nm} \equiv H_{nm}/E_1^{(0)} \nonumber \\
&=& \delta_{nm}\biggl\{n^2 +{\pi^2 \over 48}\biggl({\hbar \omega \over E_1^{(0)}}\biggr)^2
 \biggl(1 - {6 \over (\pi n)^2}\biggr)
  \biggr\}
\nonumber \\
&+& (1-\delta_{nm}) \biggl({\hbar \omega \over E_1^{(0)}}\biggr)^2 g_{nm},
\label{ham_ho}
\eea
where \be g_{nm} = \Big( {(-1)^{n+m}+1 \over 4} \Big) \Big({1 \over (n-m)^2} - {1
\over (n+m)^2} \Big). \label{gnm} \ee Note that the $g_{nm}$ remain of order unity close to the diagonal, but
for large $n$ the diagonal elements grow as $n^2$, so the off-diagonal elements become negligible in
comparison.

Equation~(\ref{ham_ho}) can be evaluated up to some cutoff for a given ${\hbar \omega/E_1^{(0)}}$; these form
the elements of a matrix to be fed into an eigenvalue/eigenvector solver (see for example, Numerical Recipes\cite{press}) or in
software packages such as Matlab, Mathematica, or Maple.

\begin{figure}[tp]
\begin{center}
\includegraphics[height=3.5in,width=3.0in, angle=-90]{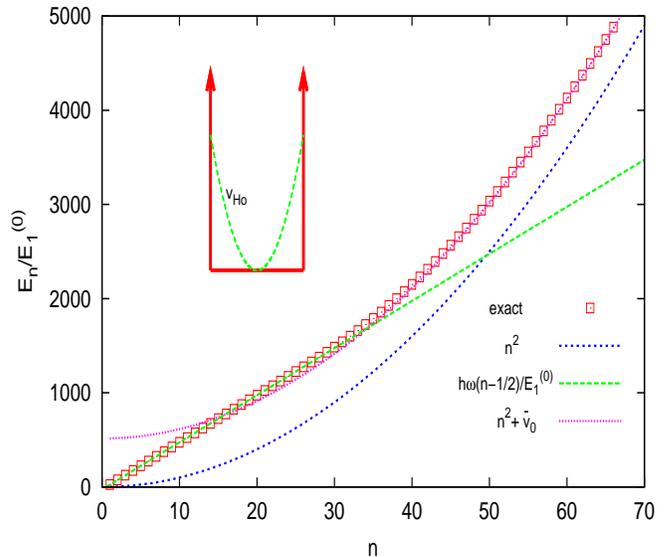} \caption{(color online)\label{fig1}
{Eigenvalues obtained by the exact diagonalization of the harmonic oscillator embedded in an
infinite square well (shown as an inset). The symbols denote the numerically converged result, obtained with a truncation
$N=400$. We also show the known analytical result for
a harmonic oscillator, $E_{\rm HO} = \hbar \omega (n - 1/2)$ (we use an enumeration that starts with one
(not zero) consistent with the enumeration of the infinite square well eigenvalues). The agreement is excellent for low-lying eigenvalues. For eigenvalues that are comparable to or
exceed the crossover energy the analytical results for the infinite square well, suitably modified to take
approximate account of the harmonic oscillator potential, are extremely accurate (dotted curve).
We used ${\hbar \omega/E_1^{(0)}} = 50$.}}
\end{center}
\end{figure}

A typical result is shown in Fig.~\ref{fig1}. The low lying states are well described by the harmonic
oscillator eigenvalues. The harmonic oscillator potential becomes truncated by the infinite square well at an
energy given by the value of the potential at $x=a$ (see Eq. \ref{ho_pot}): $E_{\rm cross}/E_1^{(0)} = (\pi^2/16)(\hbar \omega/E_1^{(0)})^2$. For this example, ${\hbar \omega/E_1^{(0)}} = 50$, so this crossover occurs at $E_{\rm
cross}/E_1^{(0)}\approx 1550$, which is approximately where the numerical results in Fig.~\ref{fig1} begin to
deviate from the analytical harmonic oscillator result. For higher energies the numerical results begin to
resemble those of the infinite square well, albeit with an added constant, $\bar{v}_0$. Inspection of
Eq.~(\ref{ham_matrix}) (see also Eq.~(\ref{ham_ho})) shows that this constant is the average of the harmonic
potential.

What happens if the harmonic potential is considerably less stiff?
If, for example, ${\hbar \omega/E_1^{(0)}} = 1$, then none of the eigenstates will describe harmonic
oscillator states; rather they all will be minor perturbations of the infinite square well eigenstates.
Nonetheless, to describe such an oscillator is still possible, simply by choosing a much wider infinite square well.

\begin{figure}[tp]
\begin{center}
\includegraphics[height=3.5in,width=3.0in, angle=-90]{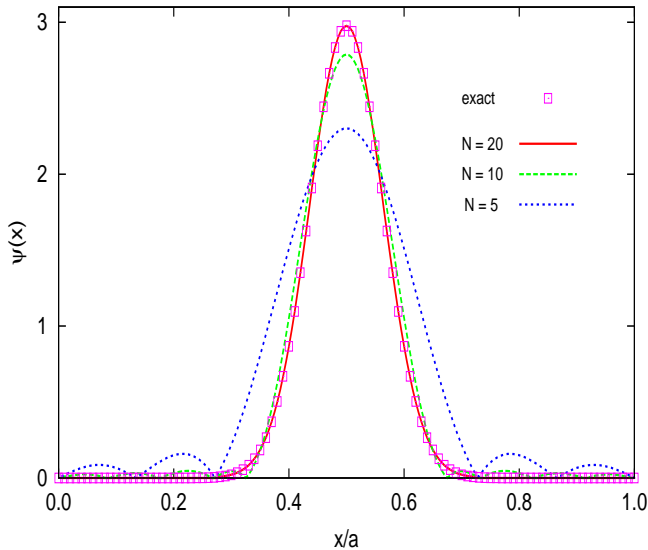} \caption{\label{fig2}(color online)
{The
ground state of the harmonic oscillator using the analytical result, Eq.~(\protect\ref{state_ho}) (symbols),
along with numerical results, obtained from matrix diagonalization, with various truncation values.
Convergence is achieved for $N=20$. We used ${\hbar \omega/E_1^{(0)}} = 50$.}}
\end{center}
\end{figure}

How well is the wave function reproduced by this truncated diagonalization technique? We use Eq.~(\ref{a2})
and obtain \be \psi(x) = \langle x|\psi \rangle = \sum_{m=1}^\infty c_m \sqrt{2 \over a} \sin{m \pi x \over
a}, \label{expansion} \ee where the $c_m$ are the components of the eigenvectors in the basis of the infinite
square well eigenfunctions. For numerical results we use Eq.~(\ref{expansion}) with the coefficients obtained
from the eigenvectors determined by the matrix diagonalization. For the harmonic oscillator the results are known
analytically. For example, in dimensionless units the ground state wave function is \be \psi_{HO}(x) =
\bigg({\pi \over 2a^2}{\hbar \omega \over E_1^{(0)}} \bigg)^{1/4} \exp \bigg[ -{\pi^2 \over 4} {\hbar \omega
\over E_1^{(0)}}\big({x \over a} - {1 \over 2} \big)^2\bigg]. \label{state_ho} \ee In Fig.~\ref{fig2} we plot
the numerical results superimposed with the analytical ones. The agreement is excellent (beyond what can be
discerned on a graph with this scale). We can also check the components $c_m$ (see Eq. (\ref{expansion})) directly,
because they are determined through diagonalization, and compare them with the expected Fourier components of
the Gaussian wave function. For example, for the ground state, these are
\be c_n = \begin{cases} i^{(n-1)}\bigl({32 \over \pi}{E_1^{(0)} \over \hbar \omega} \bigr)\exp{[-n^2 E_1^{(0)}/\hbar \omega]} &
\text{for $n$ odd,} \\ 0 & \text{for $n$ even.}
\end{cases}
\label{components} \ee

The exponential decay with quantum number $n$ helps to explain why so few components are required
to obtain an accurate representation of the Gaussian ground state wave function. When we checked the
numerically determined coefficients against those given by Eq. (\ref{components})
(not shown), we found excellent agreement.

The important point is that low lying states for any potential can be obtained to any desired precision by
setting up the matrix in Eq.~(\ref{ham_matrix}). Analytical evaluations of the integrals are not even
required.\cite{remark1} For the harmonic oscillator this method requires the least amount of mathematics. More
importantly, it serves to illustrate in a very concrete and elementary way the matrix formalism that is often
presented only abstractly in textbooks.

\section{Results for Some standard one-dimensional potentials}

There are a variety of even simpler one-dimensional potentials, and this section illustrates how results from
matrix mechanics can make contact with known analytical solutions.

Consider the potential of the form, \be V_0(x) = \begin{cases} 0 & \text{if $0 < x < b_0$,} \\ V_0 & \text{if
$b_0 < x < b_1$,} \\ 0 & \text{if $b_1<x<a$,} \\ \infty & \text{otherwise}.
\end{cases}
\label{box} \ee This potential represents a generic barrier (well) for $V_0$ greater (less) than zero.
This case can be solved
analytically although the algebra is tedious. The units are as before, and we let $\rho_0 \equiv b_0/a$,
$\rho_1 \equiv b_1/a$, and $\rho = \rho_1 - \rho_0$.

Evaluation of Eq.~(\ref{ham_matrix}) yields
\bea
&&h_{nm} \equiv H_{nm}/E_1^{(0)} \nonumber \\
&=& \delta_{nm}\biggl\{n^2 +v_0 \bigl[\rho - {\sin{2\pi n\rho_1} - \sin{2\pi n\rho_0} \over 2 \pi n}\bigr]\biggr\}
\nonumber \\
&+& v_0(1-\delta_{nm})\biggl\{ {\sin{(n-m)\pi \rho_1} - \sin{(n-m)\pi \rho_0} \over \pi (n-m)}
\nonumber \\
&& \phantom{aaaaaa} -{\sin{(n+m)\pi \rho_1} - \sin{(n+m)\pi \rho_0} \over \pi (n+m)} \biggr\}
\biggr\}.
\label{ham1}
\eea
As with the harmonic oscillator, Eq.~(\ref{ham1}) can be evaluated up to some cutoff, and the resulting matrix problem represented by Eq. (\ref{a4}) can be substituted into an
eigenvalue/eigenvector solver. We now consider some special cases.

\subsection{\label{sec:prev}Step potential}

\begin{figure}[tp]
\begin{center}
\includegraphics[height=3.4in,width=3.0in, angle =-90]{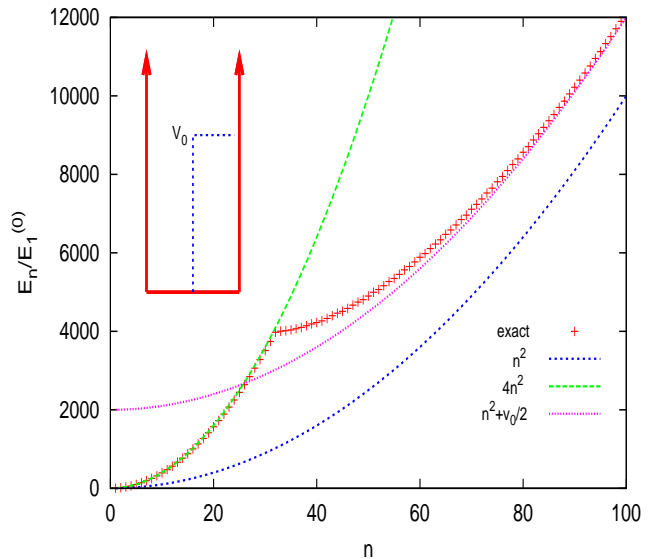} \caption{\label{fig3}(color online)
{Eigenvalues plotted as a function of quantum number $n$ for the potential shown in the inset. Symbols denote
numerically determined eigenvalues (for a matrix truncation $N=400$, but a much smaller value yields identical
results). The dashed curve illustrates the analytical result expected for an infinite square well of width
$a/2$. The short-dashed curve shows the same result for a well of width $a$. These analytical results agree with the numerical results for low and high eigenvalues, respectively. Agreement for high
eigenvalues improves significantly when a shift of $v_0/2$ (dotted curve) is included to account approximately
(as far as the high energy states are concerned) for the presence of the step potential. We used $v_0 = 4000$;
the value of $N$ is not important if it exceeds the maximum quantum number displayed by about an additional 10. }}
\end{center}
\end{figure}

One of the simplest potentials is that of a potential barrier as depicted in the insert of Fig.~\ref{fig3}.
The expression in Eq.~(\ref{ham1}) applies, with $b_0 = a/2$ and $b_1 = a$. Figure~\ref{fig3} shows the
eigenvalues as a function of the quantum number $n$. The symbols show the numerically obtained eigenvalues,
and the dashed curve shows the result expected from
Eq.~(\ref{infinite_square_well_energies}), with width $a/2$ -- hence the extra factor of 4. For low-lying
eigenvalues the agreement is very good. In other words, the particle finds itself in an essentially infinite square well
of width $a/2$. However, for sufficiently high energy states the particle should act like it is in an infinite square
well of width $a$. We have also plotted the result expected for an infinite square well of width $a$, given by
the short-dashed curve. Some accommodation should be made for the fact that a step in the potential exists at
low energies. (Imagine a state with (dimensionless) energy more than 100\,000 on the scale of Fig.~\ref{fig3}
--- the step with height $v_0 = 4000$ would appear to be a small blip.) The average of this potential is just the product
of the height and the width, so the dotted curve corresponds to the dashed
curve shifted by $v_0/2$. There are a number of states whose energies are close to $v_0$, where the
eigenvalues are not simply related to those of an infinite square well. Beyond quantum number $n=80$ the
eigenvalues, with a simple accounting for the step $v_0/2$, agree with those of an infinite square well. This
example also illustrates how the numerical solution converges to known analytical results at low and high
energies.

\subsection{Various square wells}

Consider a square well with arbitrary width less than $a$, with $V_0 < 0$. This interaction can be solved analytically, although the algebra is
tedious. For $V_0 <0$ there will exist states that are bound within the inner finite well. Insofar as these
states don't ``feel'' the walls of the infinite square well at $x/a =0,1$,
these energies and their eigenfunctions should be given by the simple analytical solution found in textbooks.
The ground state is obtained by finding the lowest positive solution to the equation,\cite{griffiths} \be
\tan{\pi u \over 2} = \sqrt{\bigg({u_0 \over u}\bigg) - 1}, \label{trans} \ee where $b \equiv b_1 - b_0$, $u
\equiv \dfrac{b}{a}\sqrt{E - V_0 \over E_1^{(0)}}$, and $u_0 \equiv \dfrac{b}{a} \sqrt{- {V_0 \over
E_1^{(0)}}}$ (for $V_0<0$ and $V_0<E<0$). In the dimensionless units used in Sec.~\ref{sec:prev}, $u = \rho
\sqrt{e - v_0}$, and similarly for $u_0$.

\begin{figure}[tp]
\begin{center}
\includegraphics[height=3.7in,width=3.7in, angle =-90]{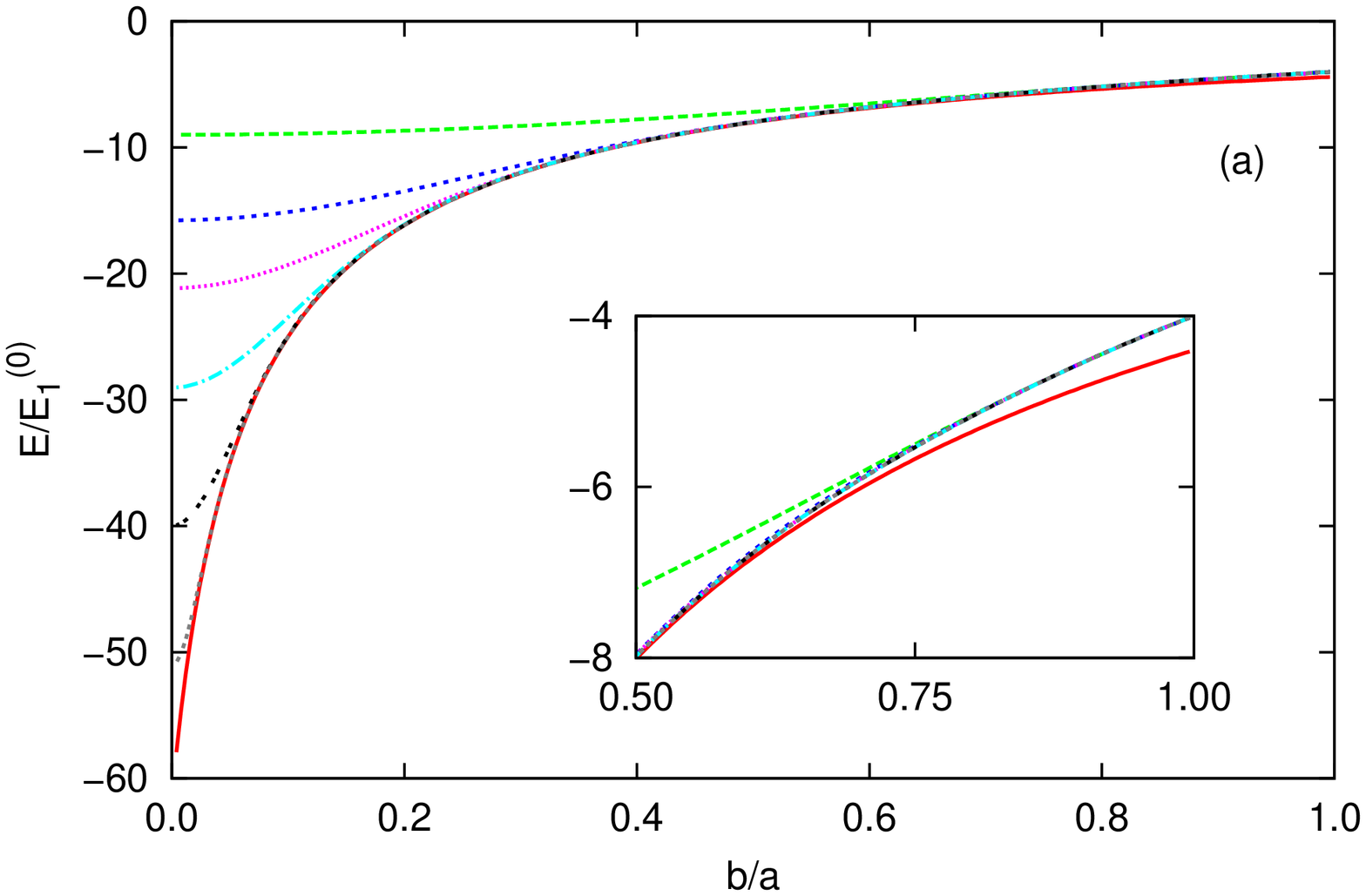}
\includegraphics[height=3.0in,width=3.0in, angle = -90]{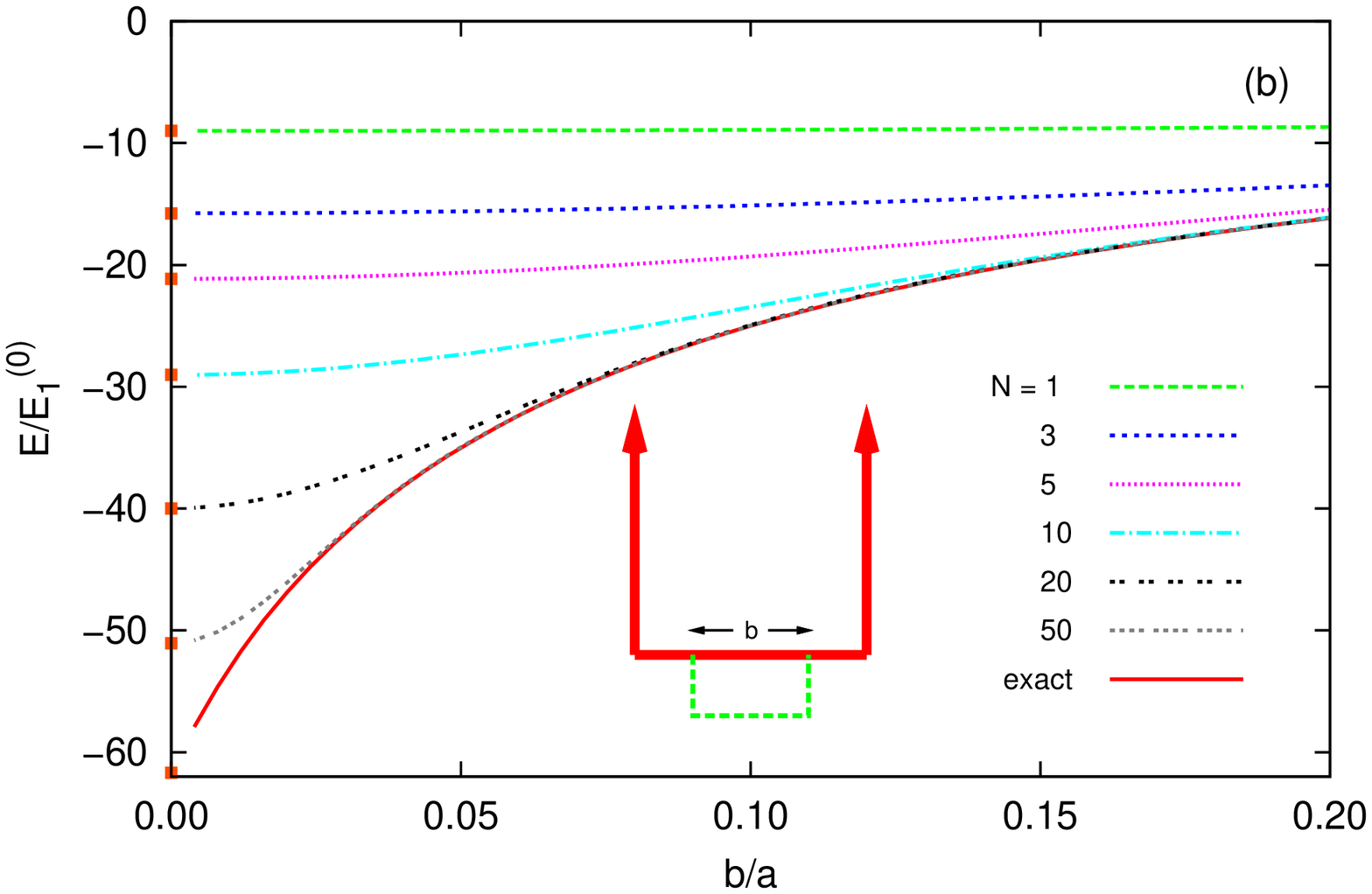} \caption{\label{fig4}(color online)
{Ground state energy as a function of finite square well width. The finite square well is imbedded
symmetrically within the infinite square well, as shown in the insert. We use $V_{00} = -5 E_1^{(0)}$, and
vary $V_0 = V_{00} a/b$ as we vary $b$. Part (a) shows that convergence as a function of truncation (see
legend in part (b)) is slow for small $b/a$. The inset shows the situation at large $b/a$, where the results
have converged, but not to the analytical result for a square well in an infinite space, but rather, as
expected, to the trivial result expected for a slightly deeper infinite square well. In part (b) we expand the low $b/a$ portion of (a), and also show analytical results for the
$\delta$-function potential, shown as symbols (see Appendix). Convergence is slow because of the need to
capture very small length scales.}}
\end{center}
\end{figure}

For the finite square well problem analytical results are provided by the solution to
Eq.~(\ref{trans}),\cite{remark0} and numerical results are obtained by diagonalizing the $N\times N$ matrix
with the elements given by Eq.~(\ref{ham1}). Specifically, we use $b_0 = a/2 - b/2$, $b_1 = a/2 + b/2$, and
$V_0 = V_{00}a/b$. For a fixed value of $V_{00}$, this choice of potential (see Fig.~\ref{fig4}) allows a study of the eigenenergies
and eigenstates of square wells with width $b$ ($0 < b < a$), keeping the product $b V_0$ fixed. The limit $b
\rightarrow 0$ describes an attractive delta function potential in the middle of the infinite square well,
which is treated analytically in the Appendix. The other limit, $b \rightarrow a$, lowers the baseline of the
infinite square well by an amount $V_0$, so that all the eigenvalues are shifted: $E_n/E_1^{(0)} \rightarrow
n^2 - V_{00}/E_1^{(0)}$ (for $b=a$ only).

Results for the ground state energy as a function of the finite square well width $b$ are shown in
Fig.~\ref{fig4}, for various values of the cutoff $N$ (see Fig.~\ref{fig4}(b)). The results are insensitive to
the cutoff for $b \approx a$, while for low values of $b$, there is considerable variation. The inset in
Fig.~\ref{fig4}(a) clarifies the situation for $b \approx a$. The result converges very rapidly (just the
diagonal element is exactly correct for $b=a$). Even the $3 \times 3$ matrix yields accurate ground state
energy values all the way down to $b \approx 0.5 a$, and as the full frame of Fig.~\ref{fig4}(a) shows,
finite $N$ results ``peel off'' as the width decreases. Note that the analytical result given by
Eq.~(\ref{trans}) differs from the numerical result near $b = a$ (see inset), because in this regime the two
approaches solve different problems. The analytical result is for a finite square well of width $b$ with no
background potential, and the numerical result includes the infinite square well in addition.\cite{remark0}
Not surprisingly, the latter case reproduces the analytical shift noted earlier from $E_0/E_1^{(0)} = 1$ to
$E_0/E_1^{(0)} = 1 - 5 = -4$ (with $V_{00} = -5 E_1^{(0)}$).

\begin{figure}[tp]
\begin{center}
\includegraphics[height=3.4in,width=3.0in, angle =-90]{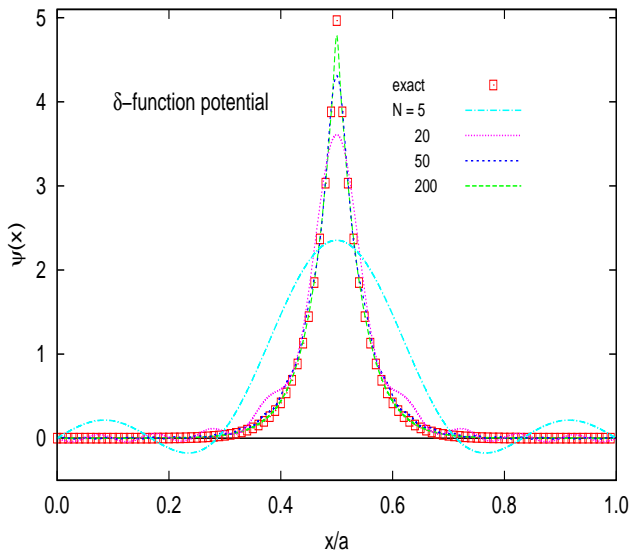} \caption{\label{fig5}(color online)
{Ground state energy wave function for a $\delta$-function potential, as calculated by diagonalization of
matrices truncated at $N=5, 20$, 50, and 200. The exact result (see Appendix) is given by symbols. There is a
cusp at the center, which requires smaller and smaller wavelength components to capture exactly. We used a
$\delta$-function with strength $b V_{00} = -5a E_1^{(0)}$, corresponding to the limiting case studied in
Fig.~\ref{fig4}.}}
\end{center}
\end{figure}

Figure~\ref{fig4}(b) clarifies the behavior in the very narrow well regime. Results for the delta function
potential are given with symbols along the $b=0$ line. The lowest curve is given by the solution of a simple equation (see Eq. (\ref{trans2}) in the Appendix). There is a considerable variation
with the cutoff $N$ in the low $b/a$ regime. For example, even with a cutoff of $N = 100$,
meaning that unperturbed states with energy as high as 10000 are included, the result at $b=0$ is off from the
$N\rightarrow \infty$ result by about 10\%. The reason is that as the finite well becomes narrower and the
wave function becomes more confined, the latter contains components with very small (of order $b$) spatial
variations. These can only be described by basis states with a similar strong variation, and such states are
to be found only among the very high energy states of the infinite square well.\cite{blume}  We use
Eq.~(\ref{expansion}) and show in Fig.~\ref{fig5} the wave functions obtained numerically for the case of a
delta function potential as we vary the cutoff of the size of the matrix retained. It is clear that for low
cutoffs, only variations on a large length scale can be captured. Even for the highest cutoff used here,
$N=200$, the numerical result still misses (though only slightly) the cusp at the centre of the figure.
The same kind of plot can be examined for square well potentials of varying width, so that
one can see directly that the matrix size required for convergence increases with narrower well width $b$
(as already seen through the energy in Fig.~\ref{fig4}).

\section{Summary}

The purpose of this paper is to provide some examples accessible to undergraduate students learning quantum
mechanics for the first time that utilize matrix diagonalization.\cite{remark2} Our goal is  to provide a
context within which a simple basis set can be used. The most straightforward case in one dimension is the
infinite square well. We worked through some examples that are generally used in an introductory course on
quantum mechanics,\cite{griffiths} with an emphasis on the harmonic oscillator. We showed that by embedding a
(truncated) harmonic oscillator potential in an infinite square well, we can obtain extremely accurate
solutions.

We also illustrated how to calculate wave functions  and demonstrated that potentials  with short length
scales often require a large number of basis states for an accurate description.

The empowering aspect of this approach is that students can tackle bound state problems for any potential. For
example, it is simple to find the low-lying states for a double well potential. The potential does not even
have to be defined analytically.\cite{remark1}

\begin{acknowledgments}

This work was supported in part by the Natural Sciences and Engineering Research Council of Canada (NSERC), by
ICORE (Alberta), and by the Canadian Institute for Advanced Research (CIfAR). The author thanks Cindy Blois
and James Day for suggestions to improve the presentation in the manuscript.

\end{acknowledgments}

\appendix

\section{Delta function potential well}

Consider a delta function potential well with strength $\alpha = V_0 b$ located at position $a_0$, inside an
infinite square well between $x=0$ and $x=a$, that is,
$$
V(x) = \begin{cases} -b V_0 \delta(x-a_0) & (0 < x < a) \\ \infty & \text{otherwise,}
\end{cases}
\label{delta}
$$
where $0 < a_0 < a$ (the symmetric case is for $a_0 = a/2$). The length scale $b$ is introduced to make
contact with the finite square well potential discussed in Sec.~III. The application of matching criteria
yields the solution (we confine our discussion to the symmetric case only):
\begin{equation}
\psi(x) = A
\begin{cases}
\sinh(\kappa x)/\sinh(\kappa a/2) & (x < a/2) \\ \sinh(\kappa (a-x))/\sinh(\kappa a/2) & (x > a/2).
\end{cases}
\label{wave_delta}
\end{equation}
The bound state energy $E$ is given by the solution of \be \tanh(\kappa a/2) = \dfrac{2\kappa}{\lambda},
\label{trans2} \ee where $\kappa \equiv \sqrt{-\dfrac{2m}{\hbar^2} E}$ and $\lambda \equiv
-\dfrac{2m}{\hbar^2}V_0 b$. Equation~\eqref{trans2} must be solved numerically. The normalization constant is
given by \be A = \sqrt{\frac{2}{a}} {1 \over \sqrt{1 - \Big(\dfrac{\lambda} {2 \kappa} \Big)^2 +
\dfrac{\lambda}{\kappa^2 a}}}. \label{norm} \ee The more familiar solution is obtained by allowing
$a\rightarrow \infty$, in which case $\kappa = \lambda/2$, and a solution always exists. In this problem, with a finite infinite square well width $a$, a solution exists only if \be {V_0 \over E_1^{(0)}} > {4 \over \pi^2}{a
\over b}. \label{condition} \ee A challenging exercise for students is to check how well Eq.~(\ref{condition})
holds for a finite square well of non-zero width $b$. Results for non-symmetrically placed delta functions can
also be readily obtained, but are not as concise, and yield a condition for existence of a bound state which
is more stringent than Eq.~(\ref{condition}).


\begin{thebibliography}{1}
\bibitem{remark1} At least one of my students simply programmed Matlab to evaluate Eq.~(\ref{ham_matrix})
    numerically for an arbitrary potential. Though not to my personal taste, many students find such a
    numerical evaluation much easier than slogging through analytical integrals. The numerical approach allows
    one to solve for bound states of any one-dimensional short range potential.
\bibitem{remark0} The analytical solution given by Eq.~(\ref{trans}) provides the solution for a finite square
    well in an infinite space. Results will therefore differ from the numerical results as the width of the
    finite square well approaches the width of the infinite square well. The Appendix provides an analytical
    solution for the $\delta$-function potential in an infinite square well, which differs from the textbook
    solution of an infinite square well in infinite space.
\bibitem{blume} These difficulties are well known. For example, bosons in a harmonic trap experience a short
    range boson-boson interaction. If we try to expand the many-particle wave function in terms of the
    eigenstates of the harmonic oscillator potential, convergence difficulties arise because of the disparity
    in length scales. Our problem illustrates this difficulty at the single particle level. I am indebted to
    Doerte Blume and Pierre-Nicholas Roy for discussions on this point.
\end{thebibliography}
\end{document}